\documentclass[twocolumn,a4paper]{jpsj3}
\usepackage{txfonts}
\usepackage{bm}
\usepackage{graphicx}

\title{Exciton-Mott Physics in Two-Dimensional Electron-Hole Systems:\\ Phase Diagram and Single-Particle Spectra}

\author{\name{Kenichi \surname{Asano}}\thanks{E-mail: asano@phys.sci.osaka-u.ac.jp} and \name{Takuya \surname{Yoshioka}}}
\inst{Department of Physics, Osaka University, Toyonaka, Osaka, 560-0043, Japan}

\abst{Exciton Mott physics in two-dimensional electron-hole (e-h) systems is studied in the quasiequilibrium, which is the crossovers or phase transitions between the insulating exciton gas and the metallic e-h plasma.
By developing a self-consistent screened $T$-matrix approximation, we succeed in obtaining the ``global'' phase diagram on the plane of the e-h density and the temperature as a contour plot of the exciton ionization ratio.
The detailed features of the exciton-Mott crossover at high temperature are figured out beyond the conventionally used concept of the Mott density.
At low temperature, we find not only the region unstable toward the inhomogeneity but the pure Mott transition point characterized by the discontinuity in the ionization ratio.
The single particle spectra also exhibit interesting features reflecting the excitonic correlations.}

\kword{exciton, electron-hole plasma, Mott transition, exciton ionization ratio, optical gain, self-consistent screened T-matrix approximation}

\begin{document}
\maketitle

\section{Introduction}
Theoretical description of the phase diagram of the electron-hole (e-h) system with long-range Coulomb interaction is a long-standing problem. 
It is expected to show a variety of physics between the insulating exciton gas and the metallic e-h plasma phases: the exciton-Mott crossovers\cite{Mott1961,Zimmermann1978,Schmitt-Rink1982,Lowenau1982,Lindberg1994,Zimmermann1985,Koch2003,Binder2009}, the gas-liquid transitions\cite{Keldish1968,Combescot1972,Brinkmann1973,Vashishta1973,Silver1973,Mahler1975,Sander1976,Haug1976,Kremp1975}, the discontinuous changes in exciton ionization ratio\cite{
Stoltz1979,Pavesi1989,Snoke1995,Arndt1996,Semkat2009}, and so on.
The e-h pair condensations are also predicted to take place at extremely low temperatures\cite{Cloizeaux1965,Keldysh1965,Keldysh1968,Jerome1967,Comte1982,Nozieres1985}.
The exciton-Mott crossover is particularly important from the viewpoint of the application to the semiconductor laser devices, since it roughly corresponds to the appearance of optical gain -- the source of the lasing. 
\par
One of the ideal stages for this study of the exciton-Mott physics is the {\it e-h bilayers}, where the electrons and holes are confined in the spatially separated planes, respectively.
They exhibit abrupt changes in the peak width of the photoluminescence (PL) spectra\cite{Timofeev2000,Stern2008}, which is never observed in the single-layer e-h systems\cite{Kappei2005,Deveaud2005}.
At the lower temperatures, indications of the e-h pair condensation are also reported\cite{Butov1994,Butov1998,Butov2002}. 
When the inter-layer distance is not large, the e-h spatial separation mainly works to extend the e-h recombination time, promoting the electrons and holes to reach a quasi-thermal equilibrium\cite{Fukuzawa1990,Golub1990,Kash1991,Damen1990,Robart1995}, as seen in the indirect semiconductors.
Therefore, some essential features of the e-h bilayer systems can be captured theoretically in the {\it single-layer e-h systems} with an artificially long e-h recombination time.
\par
One of the characters of the 2D systems is the stepwise density of states (DOS) of the non-interacting carriers, which brings two competing effects on the exciton-Mott physics.
In the lowest e-h (pair) density limit, the exciton is more stabilized in 2D systems than in the bulk (three-dimensional) cases\cite{Miller1981,Tarucha1984,Lee1979,Bastard1982,Greene1983,Greene1984,Miller1985}.
In fact, the enhancement of the density of states near the band edge results in the larger number of the low-lying energy states available for the exciton formation.
In particular, even an infinitesimally small attractive e-h interaction is strong enough to form an exciton.
The excitons might be further self-stabilized, since the plasma screening effect is suppressed as the exciton is formed. 
By contrast, increasing DOS at the band-edge also enhances the Pauli-blocking (or phase-space-filling) effect, which destabilizes the excitonic bound states at finite e-h densities.
This also means that the population inversion of e-h plasma is more easily realized than in the bulk systems, which is the reason why the 2D e-h systems are considered to be a candidate of the low-threshold laser\cite{Arakawa1982}.
\par
Three features of the exciton-Mott physics should be distinguished.
The first one is the {\it exciton-Mott crossover} between exciton gas and e-h plasma seen at high temperatures\cite{Andryushyn1976,Schmitt-Rink1984,Haug1985,Kleinman1985,Schmitt-Rink1985,Schmitt-Rink1986,Ell1989,Bongiovanni1989,Sarma1989,Ell1990,Schuster1992,Ryan1993,Ando2004,Pereira1998,Portnoi1999}. 
Experimentally, this crossover is observed as the drastic but continuous reduction of the excitonic resonance in the interband absorption-gain and in the photoluminescence spectra.
\par
The second feature is the first-order {\it gas-liquid transition} accompanied by the {\it phase coexistence}\cite{Kuramoto1974,Ishihara1982,Kleinman1986}, which is actually observed in the bulk indirect semiconductors\cite{Hensel1977}, and also in the type-II structures in direct semiconductors\cite{Kalt1992}.
In theories assuming the homogeneity of the system, this phase transition is often seen as the divergence of the isothermal compressibility which implies the instability toward the inhomogeneity.
The coexisting region can be determined by means of the Maxwell's equal-area construction of the isothermal curve of the chemical potential of the e-h pair (e-h chemical potential).
\par
The last one is the discontinuous changes in the exciton ionization ratio\cite{Rienholz2002,Leon2003,Portnoi2004,Manzke2012} -- the portion of the carriers moving nearly freely to the total electrons and holes.
This first order transition is called {\it pure Mott transition} in the following.
The abrupt changes in the PL-peak width found in the experiments might be the sign of the phase transition of this kind.
However, theoretical understanding of this phenomenon is still far from being complete.
It is also noteworthy that this type of phase transition can be regarded as the direct analogue of the Mott transition discussed in the half-filled Hubbard model, where the double occupancy, or equivalently, the fraction of the on-site exciton -- the ratio of the number of the on-site pair of the up-spin electron and the vacancy of the down-spin electron to the number of the sites -- changes discontinuously depending on the on-site Coulomb interaction.
\par
Ever since Mott himself argued\cite{Mott1961}, the exciton-Mott physics has been considered in terms of the {\it Mott density}, at which the effective exciton binding energy vanishes.
This concept is figured out most clearly by means of the so-called semiconductor-Bloch equation (SBE)\cite{Lindberg1994,Haug1985,Schmitt-Rink1985,Schmitt-Rink1986,Ell1989,Ell1990}
, a theory assuming a single e-h pair embedded in the completely ionized e-h plasma, neglecting the finite carrier lifetime induced by the inter-carrier scattering.
More specifically, the self-energies are evaluated with the quasistatic screened Hartree-Fock approximation (SHFA), and the excitonic correlation is taken into account via the screened e-h $T$-matrix.
As the e-h density increases, both the Coulomb-hole and the screened exchange self-energies reduce the effective band-gap energy, which is referred to as the band-gap-renormalization (BGR) effect.
Meanwhile, the excitonic bound state energy remains almost unchanged since it is charge neutral, and thus is hardly affected by the surrounding e-h background.
At the Mott density, this exciton level merges into the continuum states lying above the renormalized band gap.
\par
In order to go beyond this Mott-density picture and to obtain the unified and detailed view of the exciton-Mott physics, we need to deal with the mixture of the exciton gas and the e-h plasma.
The definition of the excitonic bound state should also be reconsidered, because the quasielectrons, quasiholes, and thus their excitonic bound states have finite lifetimes due to the inter-carrier scatterings.

For this purpose, the self-consistent screened $T$-matrix approximation (SSTA) serves as an efficient tool\cite{Stoltz1979,Zimmermann1985,Binder2009,Semkat2009,Pereira1998,Manzke2012}. 
In this approximation, the presence of the excitonic bound states is properly reflected to the single-particle spectra.
It also provides us with a reasonable definition of the exciton ionization ratio.
In Refs.~\citen{yoshioka2011} and \citen{yoshioka2012}, the authors have previously developed a new version of SSTA, in which the $T$-matrices, the screening parameter, and the self-energies are evaluated so as to be consistent with each other.
The efficiency of this approximation is further demonstrated by the application to the quasi-one-dimensional (quasi-1D) e-h systems, which successfully illustrates the exciton Mott physics over the wide range of the e-h density and the temperature.
In the present paper, this method is generalized to 2D e-h systems by adopting the partial-wave expansion technique.
\par
The paper is organized as follows. In \S\ref{model}, we introduce the model Hamiltonian and explain briefly our self-consistent screened $T$-matrix approximation.
We show the global phase diagram and discuss the possible phase transitions in \S\ref{results}, and finally give a summary in \S\ref{summary}.
\par
\section{Formulation}\label{model}
\subsection{Model Hamiltonian}
Our model Hamiltonian for the 2D e-h system reads
\begin{equation}
\mathcal H=\sum_{a,\bm k,\sigma}\epsilon_{a,k}c^\dagger_{a,\bm k,\sigma}c^{}_{a,\bm k,\sigma}+\frac 1{2S}\sum_{\bm q}V(q) :\rho_{\rm c,\bm q}\rho_{\rm c,-\bm q}:,
\label{eq:hamiltonian}
\end{equation}
where $S$ is the area of the system, and the pair of colons stands for the normal ordering of the creation and annihilation operators between them.
The operator, $c_{a,\bm k,\sigma}$, annihilates an electron or a hole ($a=\text{e or h}$) with the in-plane wave vector, $\bm k=(k_x,k_y)$, and the spin $\sigma=\pm 1/2$.
The charge density operator is expressed as
\begin{equation}
\rho_{{\rm c},\bm q}=\sum_{a,\bm k,\sigma}s_a c^\dagger_{a,\bm k-\bm q,\sigma}c^{}_{a,\bm k,\sigma},
\label{eq:density op}
\end{equation}
where $s_{\rm e}=-1$ and $s_{\rm h}=+1$ denote the sign of the charge of the carrier.
The energy dispersions of the noninteracting electron and hole are 
\begin{equation}
\epsilon_{a,\bm k}=\frac{E_{\rm g}}2+\frac{\hbar^2 k^2}{2m_a},
\label{eq:disp}
\end{equation}
where $m_a$ and $E_{\rm g}$ are the in-plane effective mass and the band gap, respectively. 
The Coulomb potential is Fourier transformed to 
\begin{equation}
V(q)=\int d^2\bm r\frac{e^2}{\varepsilon_{\rm b}r}e^{-i\bm q\cdot\bm r}=\frac{2\pi e^2}{\varepsilon_{\rm b} q},
\label{eq:Coulomb}
\end{equation}
with the in-plane coordinate, $\bm r=(x,y)$, the elementary charge, $e$, and the background dielectric constant, $\varepsilon_{\rm b}$.
\par
In the above model, the 2D exciton binding energy is exactly evaluated as
\begin{equation}
E_{\rm 2D}=\frac{2e^4 m_{\rm r}}{\varepsilon_{\rm b}^2\hbar^2},
\end{equation}
which is enhanced by four times that of the bulk value due to the planer confinement.
The 2D exciton Bohr radius,
\begin{equation}
a_{\rm 2D}=\frac{\hbar}{\sqrt{2m_{\rm r}E_{\rm 2D}}}=\frac{\varepsilon_{\rm b}\hbar^2}{2e^2m_{\rm r}},
\end{equation}
is shrunk to half the bulk value.
\par
With the [001] GaAs/AlGaAs QWs in mind, we adopt $m_{\rm e}=0.0665m_0$ and $m_{\rm h}=0.11m_0$ with the electron rest mass, $m_0$.
The total and the reduced masses of an e-h pair are then given as $M=m_{\rm e}+m_{\rm h}=0.177m_0$ and $m_{\rm r}=m_{\rm e}m_{\rm h}/M=0.0414m_0$, respectively.
The background dielectric constant is also assumed to be $\varepsilon_{\rm b}=13$.
In more realistic models, we also have to consider the effect of finite thickness of the QW, which reduces the exciton binding energy.
While, such effect does not affect the final results as far as all the energies and lengths are scaled by $E_{\rm 2D}$ and $a_{\rm 2D}$, respectively.
\par
\begin{figure}[tbp]
  \begin{center}
\includegraphics[width=0.47\textwidth]{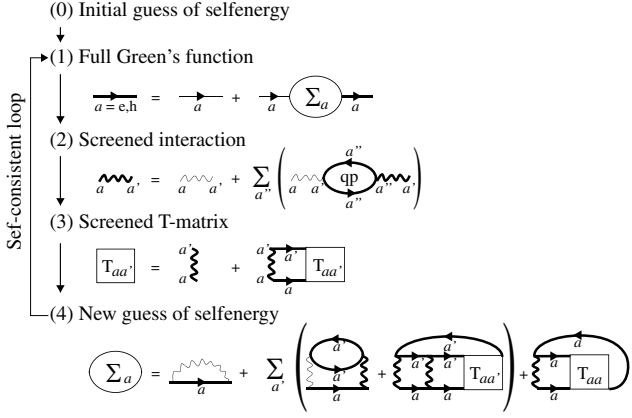}
  \end{center}
\caption{
Our self-consistent screened $T$-matrix approximation (SSTA) depicted by the Feynman diagram. 
Full and bare single-particle Green's functions are represented in thick and thin solid lines, respectively, while screened and bare Coulomb interactions are shown in thick and thin wavy lines, respectively. The polarization bubble diagram in (2) includes only the quasiparticle contributions.}
\label{fig1}
\end{figure}
\subsection{Self-consistent screened $T$-matrix approximation}
In the following, we explain the theoretical scheme of our self-consistent screened $T$-matrix approximation (SSTA), which is illustrated by the flow chart of Fig.~\ref{fig1}.
Although our SSTA is originally applied to the quasi-one dimensional (quasi-1D) e-h systems, it is straightforward to generalize it to the 2D e-h systems.
While, in order to reduce the numerical cost, we apply the partial-wave-expansion technique.
Throughout this section, we use the Planck's unit: $\hbar=1$ and $k_{\rm B}=1$.
The Green's function, the self-energies, and the $T$-matrices are the retarded ones.
\par
We start our calculation with a guess of the self-energy, $\Sigma_a(k,\omega)$, where $\omega$ is the frequency. 
The single-particle Green's function is written as
\begin{equation}
G_a(k,\omega)=\frac 1{\omega-\epsilon_{a,k}+\mu_a-\Sigma_a(k,\omega)}.
\label{G}
\end{equation}
The chemical potentials of the electron and hole, $\mu_{\rm e}$ and $\mu_{\rm h}$, are determined so as to satisfy
\begin{equation}
n=-\frac 2S\sum_{\bm k}\int \frac{d\omega}\pi f_{\rm F}(\omega){\rm Im}G_a(k,\omega)
\label{eq:e-h density}
\end{equation}
where $f_{\rm F}(\omega)=\left(e^{\omega/T}+1\right)^{-1}$ is the Fermi distribution function, and $n$ stands for the (planer) e-h density -- the number of e-h pairs per unit area.
\par
We introduce here the quasiparticle energy, $\xi_{a,k}$, as the solution of 
\begin{equation}
\xi_{a,k}=\epsilon_{a,k}-\mu_a+{\rm Re}\Sigma_a(k,\xi_{a,k}),
\label{eq:qp}
\end{equation}
which determines the renormalized band gap as
\begin{equation}
E_{\rm g}^*=\sum_a\xi_{a,k=0}.
\label{eq:BGR}
\end{equation}
and the exciton ionization ratio as
\begin{equation}
\alpha=\frac 1{nS}\sum_{a,\bm k}f_{\rm F}\left(\xi_{a,k}-\mu_a\right),
\label{eq:ion}
\end{equation}
which denotes the e-h portion behaving almost freely.
\par
In SSTA, the screening of the interaction is accounted for by the plasmon-pole approximation\cite{Haug1985,Schmitt-Rink1986,Ell1989}:
The screened Coulomb potential is evaluated as
\begin{align}
W(q)&=\lim_{\Omega\rightarrow 0}V(q)\left(1+\frac{\Omega_q^2}{\Omega^2-\overline\Omega_q^2}\right)\notag\\
&=\frac{2\pi e^2}{\varepsilon_{\rm b}}\left[q+\kappa\left[1+\frac{C\kappa q^3}{16m_{\rm r}^2\Omega_q^2}\right]^{-1}\right]^{-1}
\label{eq:screen}
\end{align}
with the 2D plasma frequency, 
\begin{equation}
\Omega_q=\left[\frac{2\pi ne^2}{\varepsilon_{\rm b} m_{\rm r}}q\right]^{1/2}.
\label{eq:plasma-freq}
\end{equation}
The effective plasmon-pole frequency is determined so as to satisfy the compressibility sum rule, and is written as
\begin{equation}
\overline\Omega_q=\left[\Omega_q^2\left(1+\frac{q}{\kappa}\right)+\frac{Cq^4}{16m_{\rm r}^2}\right]^{1/2},
\label{eq:plasmon-pole}
\end{equation}
where $\kappa$ denotes the screening parameter (the inverse of screening length), and $C$ is treated as an adjustable parameter.
In order to take into account the suppression of screening due to the exciton formation, we attribute the screening only to the quasiparticles.
In other words, the screening parameter is evaluated as
\begin{equation}
\kappa=\frac{4\pi\lambda_{\rm L}}{S}\sum_{a,\bm k}f_{\rm F}(\xi_{a,k}-\mu_a)\left[1-f_{\rm F}(\xi_{a,k}-\mu_a)\right].
\label{eq:kappa}
\end{equation}
with the so-called Landau length, $\lambda_{\rm L}=e^2/\varepsilon_{\rm b}T$.
We adopt $C=4$ in the following calculation\cite{Ell1989}, whereas the final results are insensitive to this value.
\par
Next, we consider the multiple scattering between two particles denoted by $a$ and $b$ ($={\rm e,h}$).
Their momenta change from $(\bm k,\bm Q-\bm k)$ to $(\bm k',\bm Q-\bm k')$, where $\bm Q$ is the center-of-mass momentum.
This scattering process is described most simply in the center-of-mass frame, since the initial and final pairs of momenta are given as $(\bm p,-\bm p)$ and $(\bm p',-\bm p')$, respectively.
Here, we define the relative momenta as $\bm p=\bm k-m_a\bm v_{\rm G}$ and $\bm p'=\bm k'-m_a\bm v_{\rm G}$, using the center-of-mass velocity in the original frame, $\bm v_{\rm G}=\bm Q/(m_a+m_b)$.
The $T$-matrix, $\mathcal T_{ab}(\bm p,\bm p';\bm Q,\Omega)$, for this $a$-$b$ scattering is given as the solution of the Bethe-Salpeter equation, 
\begin{align}
&\mathcal T_{ab}(\bm p,\bm p';\bm Q,\Omega)\notag\\
&=s_as_bW(|\bm p-\bm p'|)\notag\\
&\hspace{5mm}+\frac 1S\sum_{\bm p''}s_a s_b W(|\bm p-\bm p''|)\notag\\
&\hspace{15mm}\times\mathcal G^{(0)}_{ab}(\bm p'';\bm Q,\Omega)\mathcal T_{ab}(\bm p'',\bm p';\bm Q,\Omega),
\label{eq:BSE}
\end{align}
where the two-particle Green's function without vertex correction is computed as
\begin{align}
&\mathcal G^{(0)}_{ab}(\bm p;\bm Q,\Omega)=-\frac{1}{\pi}\int d\Omega'\frac{{\rm Im}\mathcal G^{(0)}_{ab}(\bm p;\bm Q,\Omega')}{\Omega-\Omega'+i\delta},\label{eq:G20}\\
&{\rm Im}\mathcal G^{(0)}_{ab}(\bm p;\bm Q,\Omega)=-\int\frac{d\omega'}{\pi}\left(1-f_{\rm F}(\omega')-f_{\rm F}(\Omega-\omega')\right)\notag\\
&\hspace{3cm}\times{\rm Im}G_a(|m_a\bm v_{\rm G}+\bm p|,\omega')\notag\\
&\hspace{3cm}\times{\rm Im}G_b(|m_b\bm v_{\rm G}-\bm p|,\Omega-\omega'),
\label{eq:ImG20}
\end{align}
with a positive infinitesimal, $\delta$.
In our calculation, the e-h $T$-matrix, $\mathcal T_{\rm eh}$ is especially important, since it describes the excitonic correlation.
\par
It is reasonable to expect that $\mathcal G^{(0)}_{ab}(\bm p;\bm Q,\Omega)$ only weakly depends on the angle, $\theta_{\bm p\bm Q}$, where $\theta_{\bm a\bm b}$ stands for the angle formed by the two vectors, $\bm a$ and $\bm b$.
This is because this Green's function becomes independent of this angle in the classical (low-density or high-temperature) limit.
Thus, we approximate it by its angle-average,
\begin{equation}
\overline{\mathcal G}^{(0)}_{ab}(p;Q,\Omega)=\int\frac{d\theta_{\bm p\bm Q}}{2\pi}\mathcal G_{ab}^{(0)}(\bm p;\bm Q,\Omega),
\label{eq:ang-av}
\end{equation}
where the integrand depends on $\theta_{\bm p\bm Q}$ via $|m_a\bm v_{\rm g}\pm \bm p|=[m_a^2v_{\rm G}^2+p^2\pm 2m_av_{\rm G}p\cos\theta_{\bm p\bm Q}]^{1/2}$ in Eq.~(\ref{eq:ImG20}).
It is also noteworthy that this approximation becomes exact at $\bm Q=0$ even at high e-h densities.
Then, the $T$-matrix becomes independent of the direction of the center-of-mass momentum, $\bm Q$, and its angle-dependence comes only from the scattering angle in the center-of-mass frame, $\theta_{\bm p\bm p'}$.
The Bethe-Salpeter equation is rewritten as
\begin{align}
&\mathcal T_{ab}(m,p,p';Q,\Omega)\notag\\
&=s_a s_bW(m,p,p')\notag\\
&\hspace{5mm}+\int_0^\infty\frac{dp''}{2\pi} p''s_a s_b W(m,p,p'')\nonumber\\
&\hspace{15mm}\times\overline{\mathcal G}^{(0)}_{ab}(p'';Q,\Omega)T_{ab}(m,p'',p';Q,\Omega),
\label{eq:BSE2}
\end{align}
where the partial-wave expansions of the screened potential and the $T$-matrix are defined as
\begin{align}
W(|\bm p-\bm p'|)&=\sum_m W(m,p,p')e^{im\theta_{\bm p\bm p'}},\\
\mathcal T_{ab}(\bm p,\bm p';\bm Q,\Omega)&=\sum_m \mathcal T_{ab}(m,p,p';Q,\Omega)e^{im\theta_{\bm p\bm p'}},
\label{eq:partial-wave}
\end{align}
with the integer, $m$.
In the following calculation, we keep only the components of $|m|\le 3$.
\par
The new candidate of the self-energy is evaluated in the spectral representation:
\begin{equation}
\Sigma_a(k,\omega)=\Sigma_a^{\rm (HF)}(k)-\int \frac{d\omega'}\pi\frac{{\rm Im}\Sigma_a(k,\omega')}{\omega-\omega'+i\delta}.
\label{eq:self}
\end{equation}
The former $\omega$-independent term,
\begin{align}
\Sigma^{\rm (HF)}(k)
&=\frac 1S\sum_{\bm k'}V(|{\bm k-\bm k'}|)\int\frac{d\omega}\pi{\rm Im}G_a(k',\omega)\notag\\
&=\int\frac{dk'}{2\pi}k'V(m=0,k,k')\int\frac{d\omega}\pi{\rm Im}G_a(k',\omega),
\label{eq:self-HF}
\end{align}
denotes the Hartree-Fock contribution, where the partial-wave expansion of the unscreened potential, $V$, is defined similarly to that of the screened potential, $W$.
\par
The latter term of Eq.~(\ref{eq:self}) expresses the correlation effects, and its imaginary part splits into four terms:
\begin{align}
{\rm Im}\Sigma_a=&{\rm Im}\left[\Sigma^{\rm (MW)}_a-\Sigma^{\rm (L2)}_a+\Sigma^{\rm (L)}_a+\Sigma^{\rm (L')}_a\right].
\label{eq:Imself}
\end{align}
Here, $\Sigma^{\rm (MW)}_a$ and $\Sigma^{\rm (L2)}_a$ stand for the Montroll-Ward and the screened-Born terms, respectively.
Their imaginary parts are explicitly written as
\begin{align}
&{\rm Im}\left[\Sigma^{\rm (MW)}_a(k,\omega)-\Sigma^{\rm (L2)}_a(k,\omega)\right]\notag\\
&=-\int\frac{d\omega'}\pi\left[f_{\rm B}(\omega+\omega')+f_{\rm F}(\omega')\right]\notag\\
&\hspace{5mm}\times\frac 1S\sum_{\bm k'}\sum_b U_{ab}^{\rm (2)}(\bm k,\bm k',\omega+\omega'){\rm Im}G_b(k',\omega'),
\label{eq:Imself2}
\end{align}
with the integral kernel,
\begin{align}
U_{ab}^{(2)}(\bm k,\bm k',\Omega)=&\frac 2S\sum_{\bm p''}\left(V(|\tilde{\bm p}-\bm p''|)-W(|\tilde{\bm p}-\bm p''|)\right)\notag\\
&\times W(|\tilde{\bm p}-\bm p''|){\rm Im}{\mathcal G}_{ab}^{(0)}(\bm p'';\tilde{\bm Q},\Omega),
\label{eq:U2}
\end{align}
where we introduce $\tilde{\bm Q}=\bm k+\bm k'$, $\tilde{\bm p}=\bm k-m_a\bm v_{\rm G}=(m_b\bm k-m_a\bm k')/(m_a+m_b)$, and the Bose distribution function, $f_{\rm B}(\Omega)=[e^{\beta\Omega}-1]^{-1}$.
Approximating ${\mathcal G}_{ab}^{(0)}$ by $\overline{\mathcal G}_{ab}^{(0)}$, we obtain the final expression:
\begin{align}
&{\rm Im}\left[\Sigma^{\rm (MW)}_a(k,\omega)-\Sigma^{\rm (L2)}_a(k,\omega)\right]\notag\\
&=-\int\frac{d\omega'}\pi\left[f_{\rm B}(\omega+\omega')+f_{\rm F}(\omega')\right]\notag\\
&\hspace{5mm}\times\sum_b\int_0^\infty\frac{dk'}{2\pi}k'\;\overline{U}_{ab}^{(2)}(k,k',\omega+\omega'){\rm Im}G_b(k',\omega'),
\label{eq:Imself2a}
\end{align}
with the kernel averaged with respect to $\theta_{\bm k\bm k'}$,
\begin{align}
\overline{U}_{ab}^{(2)}(k,k',\Omega)
&=2\int_0^{2\pi}\frac{d\theta_{\bm k\bm k'}}{2\pi}\int_0^\infty\frac{dp''}{2\pi}p''\notag\\
&\hspace{5mm}\times\sum_m\left(V(m,\tilde p,p'')-W(m,\tilde p,p'')\right)\notag\\
&\hspace{5mm}\times W(m,\tilde p,p''){\rm Im}\overline{\mathcal G}_{ab}^{(0)}(p'';\tilde Q,\Omega),
\label{eq:U2a}
\end{align}
where the integrand depends on $\theta_{\bm k\bm k'}$ through $\tilde Q=[k^2+{k'}^2+2kk'\cos\theta_{\bm k\bm k'}]^{1/2}$ and $\tilde p=[m_b^2k^2+m_a^2{k'}^2-2m_am_bkk'\cos\theta_{\bm k\bm k'}]^{1/2}/(m_a+m_b)$.
\par
The rest of the correlation is attributed to the direct and exchange contributions of $T$-matrix, $\Sigma^{\rm (L)}_a$ and $\Sigma^{\rm (L')}_a$.
Their imaginary parts, ${\rm Im}[\Sigma_a^{\rm (L)}(k,\omega)+\Sigma_a^{\rm (L')}(k,\omega)]$, have the same expression as Eq.~(\ref{eq:Imself2}), if the integral kernel is redefined as
\begin{align}
U_{ab}^{\rm (L)}(\bm k,\bm k',\Omega)
=&{\rm Im}\Bigl[2\mathcal T_{ab}(\tilde{\bm p},\tilde{\bm p};\tilde{\bm Q},\Omega)\notag\\
&\hspace{1cm}-\delta_{ab}\mathcal T_{ab}(\tilde{\bm p},-\tilde{\bm p};\tilde{\bm Q},\Omega)\Bigr].
\label{eq:UL}
\end{align}
The final expression of ${\rm Im}[\Sigma_a^{\rm (L)}(k,\omega)+\Sigma_a^{\rm (L')}(k,\omega)]$ is derived by substituting ${\mathcal G}_{ab}^{(0)}$ with $\overline{\mathcal G}_{ab}^{(0)}$.
It has the same form as Eq.~(\ref{eq:Imself2a}), if the kernel is averaged regarding $\theta_{\bm k\bm k'}$ as
\begin{align}
\overline{U}_{ab}^{\rm (L)}(k,k',\Omega)
=&\int\frac{d\theta_{\bm k\bm k'}}{2\pi}\sum_m
{\rm Im}\Bigl[2\mathcal T_{ab}(m,\tilde p,\tilde p;\tilde Q,\Omega)\notag\\
&\hspace{5mm}-\delta_{ab}(-1)^m\mathcal T_{ab}(m,\tilde p,\tilde p;\tilde Q,\Omega)\Bigr].
\label{eq:ULa}
\end{align}
\par
By iterating the above mentioned updates of the self-energies up to their convergence, the $T$-matrices, the screening parameter, and the self-energies (single-particle Green's function)  can be determined consistently with each other.
It is noteworthy here that the presence of the excitonic bound state is properly taken into account in our single-particle Green's functions, since the direct contribution of the e-h $T$-matrix is included in the self-energies.
This allows us to consider how much portion of electrons and holes occupies the bound and scattering states via the single-particle spectra, and to define the exciton ionization ratio as Eq.~(\ref{eq:ion}).
\par
\subsection{Plasma-gain-onset density}
The {\it plasma-gain-onset density} is defined by the condition,
\begin{equation}
E_{\rm g}^*=\mu\label{plasma-gain-onset density}
\end{equation}
at a given temperature, $T$, where $\mu=\mu_{\rm e}+\mu_{\rm h}$ denotes the chemical potential for an e-h pair.
The population inversion of the e-h plasma of quasiparticles is formed above this density.
\par
Not only in two dimension but also in general dimensions, most of the interaction effects on the quasiparticle energies are attributable almost solely to the band gap renormalization, and their masses are nearly unchanged from $m_{\rm e}$ and $m_{\rm h}$.  
Namely, the rigid-band-shift picture holds, and the relation between the e-h density, $n$, and the e-h chemical potential measured from the renormalized band gap, $\mu-E_{\rm g}^*$, can be derived using the free-carrier theory.
As a result, the plasma-gain-onset density, $n_{\rm P}$, is well approximated by
\begin{equation}
n_{\rm P}\lambda_{\rm T}^D\sim 2\left(\frac M{m_{\rm r}}\right)^{D/4}I_{D/2-1}(0)
\label{eq:plasma-gain-onset density}
\end{equation}
where $D$ denotes the spatial dimension, and
\begin{equation}
\lambda_{\rm T}=\frac h{\sqrt{2\pi m_{\rm r}k_{\rm B}T}}.
\end{equation}
is the thermal de-Broglie length defined with the e-h reduced mass.
The complete Fermi-Dirac integral is also numerically evaluated as
\begin{equation}
 I_{D/2-1}(0)=\left\{
\begin{array}{ll}
0.604899&(D=1)\\
\ln 2=0.693147&(D=2)\\
0.765147&(D=3)
\end{array}
\right.
\end{equation}
From this result, we can see that the plasma-gain-onset density defines a classical-quantum crossover.
It is determined only by the Pauli-blocking effect (Fermi statistics of electrons and holes), and is independent of the manybody effects.
\par
\subsection{Mott Density}\label{sec:Mott density}
In many previous theoretical studies, the exciton Mott physics is discussed in terms of the {\it Mott density}, which corresponds to the dissociation of the quasiexciton -- the bound state of the quasi-e-h pair.
In our SSTA, the concept of Mott density becomes somewhat ambiguous, because the quasielectrons and quasiholes acquire the finite lifetimes.
Still, in the low-dimensional e-h systems, we can derive analytically the approximation for the Mott density on the basis of a simple argument\cite{yoshioka2011,yoshioka2012}.
\par
For a moment, let us neglect the finite-life-time effects on the quasielectrons and quasiholes.
If the ground eigenenergy of the quasi-e-h pair is lower than the renormalized band gap energy, $E_{\rm g}^*$, it is identified as the quasiexciton energy, $E_{\rm X}^*$.
Then, the Mott density, $n_{\rm M}$, is defined as the vanishing of the effective binding energy, 
\begin{equation}
E_{\rm g}^*-E^*_{\rm X}\rightarrow 0
\label{eq:Mott density}
\end{equation}
at a given temperature, $T$.
Namely, the quasiexciton is formed at $n<n_{\rm M}$, and is dissociated at $n>n_{\rm M}$.
\par
Now, let us compare the Mott density, $n_{\rm M}$, with the plasma-gain-onset density, $n_{\rm P}$. 
Here, we exclude the possibility of the Bose-Einstein condensation (BEC) of the quasiexcitons, which implies that the quasiexciton energy, $E_{\rm X}^*$, should be less than the e-h chemical potential, $\mu$, at $n<n_{\rm M}$.
Then, we generally see $E_{\rm g}^*=E_{\rm X}^*\le \mu$ at $n=n_{\rm M}$, and obtain the inequality, $n_{\rm M}\le n_{\rm P}$.
\par
In low-dimensional (quasi-1D and 2D) e-h systems, we can further show the equality, $n_{\rm M}=n_{\rm P}$.
In fact, even an infinitesimally small attractive interaction can form a bound state of quasi-e-h pair, so that a quasiexciton is always present up to the plasma-gain-onset density.
Consequently, Eq.~(\ref{eq:plasma-gain-onset density}) leads to
\begin{equation}
n_{\rm M}=n_{\rm P}\sim 1.21\times\left(\frac M{m_{\rm r}}\right)^{1/4}\frac 1{\lambda_{\rm T}}\propto T^{1/2}, 
\end{equation}
in the quasi-1D e-h systems, and 
\begin{equation}
n_{\rm M}=n_{\rm P}\sim\sqrt{\frac{M}{m_{\rm r}}}\frac{2\ln 2}{\lambda_{\rm T}^2}\propto T,
\label{eq:mott density}
\end{equation}
in the 2D e-h systems.
Interestingly enough, not only the plasma-gain-onset density but the Mott density can be understood as a classical-quantum crossover.
\par
It is noteworthy that the equality, $n_{\rm M}=n_{\rm P}$, is characteristic of the low dimensional e-h systems.
In fact, in the bulk (3D) e-h systems, $n_{\rm M}$ and $n_{\rm P}$ independently define the two different crossovers.
At high temperature, it is known that the Mott density can be well approximated as
\begin{equation}
n_{\rm M}\sim 0.028\times \frac{k_{\rm B}T}{E_{\rm 3D}a_{\rm 3D}^3}\propto T,
\end{equation}
which is equivalent to the Mott criterion\cite{Rogers1970},
\begin{equation}
\kappa_{\rm 3D}^{\rm (DH)}a_{\rm 3D}\sim 1.19,
\end{equation}
where  the Debye-H\"uckel screening parameter, $\kappa_{\rm 3D}^{\rm (DH)}=(8\pi ne^2/\varepsilon_{\rm b}k_{\rm B}T)^{1/2}$, the exciton Bohr radius, $a_{\rm 3D}=\hbar^2\epsilon_{\rm b}/2e^2m_{\rm r}$, and the exciton binding energy, $E_{\rm 3D}=\hbar^2/2m_{\rm r}a_{\rm 3D}^2$, are evaluated in 3D.
This result shows that the Mott density is mainly due to the screening effect.
On the other hand, the plasma-gain-onset density is estimated by Eq.~(\ref{eq:plasma-gain-onset density}) as 
\begin{equation}
n_{\rm P}\sim 1.53\times\left(\frac Mm\right)^{3/4}\frac 1{\lambda_{\rm T}^3}\propto T^{3/2},
\end{equation}
and is ascribed only to the Pauli-blocking effect.
As a result, we obtain $n_{\rm M}\ll n_{\rm P}$.
\par
Strictly speaking, the case of 2D is somewhat marginal.
Although the equality, $n_{\rm M}=n_{\rm P}$, holds, the effective exciton binding energy is expected to be exponentially small near below $n_{\rm P}$.
Such a small binding energy is hardly resolved in the numerical calculation, and more importantly, is physically meaningless.
In fact, our SSTA calculation predicts that the inter-carrier scattering give considerably large broadening to the quasiparticle energies at $n\sim n_{\rm P}$.
Thus, it is more reasonable to consider that the ``realistic'' values of $n_{\rm M}$ are of the same order of but slightly smaller than $n_{\rm P}$.
\par
\section{Results}\label{results}
\subsection{Global phase diagram}
\begin{figure}[tbp]
 \begin{center}
  \includegraphics[width=0.48\textwidth]{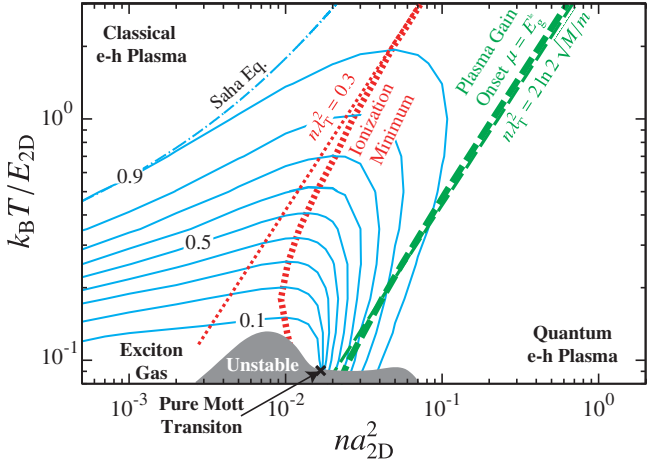}
 \end{center}
\caption{(color online) Global phase diagram of 2D e-h system depicted by contour plot of the ionization ratio, $\alpha$, on $n$-$T$ plane.
The e-h density, $n$, and the temperature, $T$, are measured in units of the square inverse of 2D exciton Bohr radius, $a^{-2}_{\rm 2D}=1.30\times 10^{12}{\rm cm^{-2}}$, and of the binding energy, $E_{\rm 2D}=139{\rm K}$, respectively.
The contour lines are shown in the solid lines, and the result by the Saha equation for $\alpha=0.9$ is also shown in the chain line.
Homogeneous thermodynamic state becomes unstable on the boundary of the gray shaded region, while $\alpha$ changes discontinuously at the cross point.
On the thick dotted line, $\alpha$ is minimized when $T$ is fixed.
At high temperature, this line asymptotically approaches a classical-quantum crossover line, $n\lambda_{\rm T}^2=0.3$, which is shown in the thin dotted line.
The thick broken line denotes the plasma-gain-onset density, which is well approximated by
the thin broken line, $n\lambda_{\rm T}^2=2\ln 2\sqrt{M/m_{\rm r}}=1.24$, defining another classical-quantum crossover.}
\label{fig2}
\end{figure}
The global phase diagram is represented by the contour plot of the ionization ratio, $\alpha$, in Fig.\ref{fig2}.
Here, the e-h density, $n$, and the temperature, $k_{\rm B}T$, are normalized by the square inverse of the 2D exciton Bohr radius, $a_{\rm 2D}^{-2}$, and by the 2D exciton binding energy, $E_{\rm 2D}$, respectively.
\par
Let us overview this phase diagram.
The e-h plasma phase (almost complete ionization) appears at low e-h density in the classical region ($n\lambda_{\rm T}^2\ll 1$), and also at high e-h density in the quantum region ($n\lambda_{\rm T}^2\gg 1$), which are considered as the weak coupling regimes.
The region of the exciton-gas (low-ionization) phase is restricted only to the low-$n$-low-$T$ region, which is the strong coupling regime.
This qualitative behavior can be understood as follows.
In the classical region, the coupling strength -- the characteristic ratio of Coulomb interaction to the kinetic energy -- is given by the so-called nonideality parameter, 
\begin{equation}
\Gamma=\frac{\lambda_{\rm L}}d=\frac{e^2/\varepsilon_{\rm b}d}{k_{\rm B}T}
\end{equation}
with the mean e-h distance, $d=1/\sqrt{\pi n}$.
Thus, the low e-h density implies the weak coupling.
In contrast, in the quantum region, the coupling strength is expressed by the $r_{\rm s}$ parameter, 
\begin{equation}
r_{\rm s}=\frac d{a_{\rm 2D}}
=\frac{e^2/\varepsilon_{\rm b}d}{\hbar^2/2m_{\rm r}d^2}
\propto\frac{e^2/\varepsilon_{\rm b}d}{E_{\rm F}}
\end{equation}
where $E_{\rm F}$ denotes the sum of the Fermi energies of the electron and the hole.
Therefore, the weak coupling regime appears at high density.
Inversely speaking, we meet the strong coupling regime only when both the e-h density and the temperature are lowered.
\par
In the classical regime ($n\lambda_{\rm T}^2\ll 1$), the ionization ratio, $\alpha$, obeys the Saha equation\cite{Saha1920,Saha1921,Kington1923},
\begin{equation}
\frac{\alpha^2}{1-\alpha}=\frac 1{n\lambda^2_{\rm T}}\exp\left(-\frac{E_{\rm 2D}}{k_{\rm B}T}\right) 
\end{equation}
which is derived from the mass-action law within the framework of the classical statistical mechanics.
We confirm that $\alpha$ is an increasing function of $T$ (thermal dissociation) and the decreasing function of $n$ (entropy dissociation). 
\par
\begin{figure}
 \begin{center}
  \includegraphics[width=0.48\textwidth]{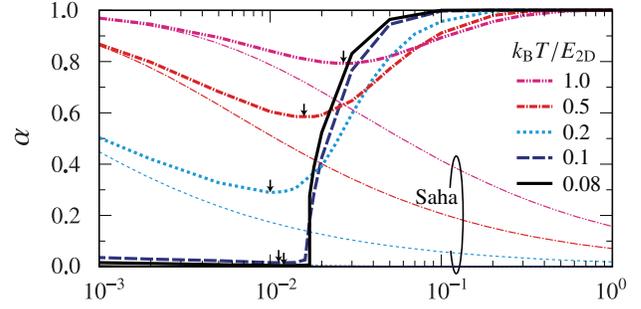}
 \end{center}
\caption{(color online) Ionization ratio, $\alpha$, is plotted in thick lines as a function of e-h density, $n$, at several values of temperature, $k_{\rm B}T/E_{\rm 2D}=0.08$,$0.1$, $0.2$, $0.5$ and $1.0$. For comparison, results of the Saha equation are also shown in thin lines. The e-h densities, $n=n_{\rm min}(T)$, which minimize $\alpha$ are indicated by arrows.}
\label{fig3} 
\end{figure}
In contrast, in the quantum regime ($n\lambda_{\rm T}^2\gg 1$), $\alpha$ increases with $n$, because the binding energy of the quasiexciton is reduced (quantum dissociation).
As a result, $\alpha$ shows the minimum at an e-h density, $n=n_{\rm min}(T)$, as indicated by arrows in Fig.~\ref{fig3}. 
The ionization ratio, $\alpha$, falls following the Saha equation at $n<n_{\rm min}(T)$, and turns to rise at $n>n_{\rm min}(T)$ as a consequence of the quantum dissociation.
At high e-h densities, the excitons are almost fully ionized, and $1-\alpha$ becomes negligibly small.
However, our calculation never gives the exciton ionization ratio exactly equal to unity, so that the contour line of $\alpha=1$, which may define a second or higher order phase transition, is absent in Fig.~\ref{fig2}.\cite{correction}
\par
In Fig.~\ref{fig2}, $n=n_{\rm min}(T)$ is represented by the thick dotted line, which most intelligibly characterizes the exciton-Mott physics, namely, the $(n,T)$ dependence of exciton ionization ratio.
As expected, this line defines a classical-quantum crossover at high temperature, and thus has the asymptote in the form of $n\lambda_{\rm T}^2=C$, which is shown in the thin dotted line.
From the numerical data, we roughly estimate the value of the constant, $C$, as $0.3$.
Whereas, at low temperatures, the line of $n=n_{\rm min}(T)$ deviates from the asymptote, and goes into the regime of quantum exciton gas.
This behavior is ascribed to the self-stabilization effect of excitons due to the screening suppression, as will be mentioned shortly.
\par
Now, we are aware of the insufficiency of the traditional argument in terms of the Mott density, $n_{\rm M}$.
As already discussed in Sec.~\ref{sec:Mott density}, there is a relationship, $n_{\rm M}\sim n_{\rm P}$ in the low-dimensional e-h systems.
Thus, the plasma-gain-onset density, $n_{\rm P}$, is plotted here in the broken line, instead of the Mott density, $n_{\rm M}$.
It can be seen that the Mott density gives only a rough description of the exciton Mott crossover, and instead, our ionization ratio gives a new insight of the Mott physics.
We can also confirm the validity of the approximation of Eq.~(\ref{eq:plasma-gain-onset density}) for the plasma-gain-onset density, $n_{\rm P}$, as show in the thin broken line.
\par
It is noteworthy that the ionization ratio in the low-$T$ region is significantly less than unity even at $n=n_{\rm P}\gtrsim n_{\rm M}$, where the binding energy of the quasiexciton vanishes.
For instance, $\alpha\lesssim 0.6$ is obtained at $n=n_{\rm P}$ at $k_{\rm B}T\sim 0.1E_{\rm 2D}$.
The reason for this behavior will be given in Sec.~\ref{sec:single-particle spectra}.
\par
At lower temperature, we found the gray region, where the homogeneous thermodynamic state becomes unstable.
Its boundary is determined by the divergence of the isothermal compressibility:
\begin{equation}
-\frac 1S\left(\frac{\partial S}{\partial p}\right)_T=\frac 1{n^2}\left(\frac{\partial n}{\partial\mu}\right)_T\rightarrow +\infty,
\end{equation}
where $p$ denotes the ``pressure'' (force per unit length) of the 2D e-h system.
The gray region shows a ``dip'' at $na_{\rm 2D}^2\sim 1.7\times 10^{-2}$.
Below the cross point near this dip, a discontinuous change in the ionization ratio -- a pure Mott transition -- is found, as seen in the data at $k_{\rm B}T/E_{\rm 2D}=0.08$ in Fig.~\ref{fig3}.
\par
We also investigate the possibility of the e-h pair condensation by the Thouless criterion which examines the divergence of the e-h pair susceptibility.
Although the susceptibility increases at low temperature, it never diverges.
In our SSTA, the pair fluctuation effect is properly taken into account, which suppresses the e-h pair condensation.
This aspect will be discussed elsewhere.
\begin{figure}[tbp]
  \begin{center}
\includegraphics[width=0.47\textwidth]{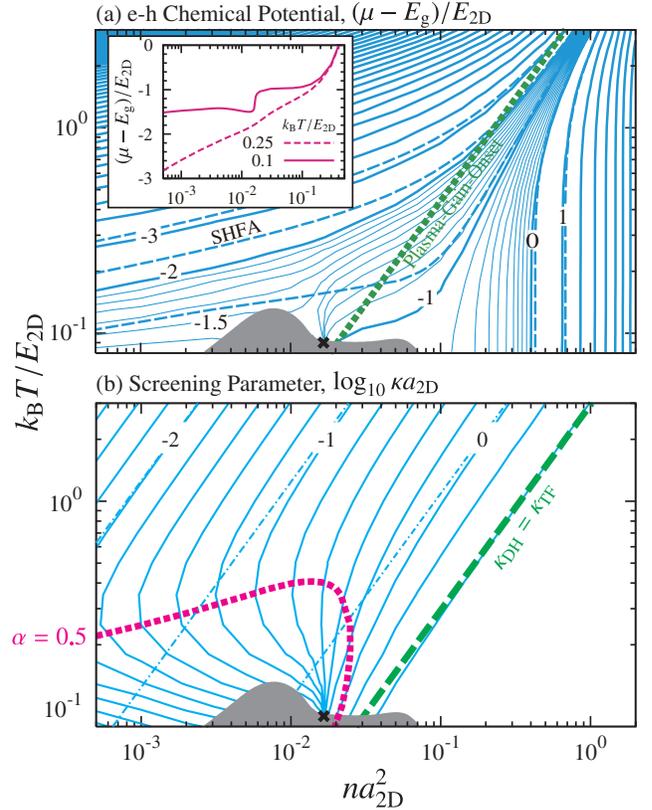}
  \end{center}
\caption{(color online) (a) Contour plot of the e-h chemical potential on $n$-$T$ plane.
Solid and broken lines show the results obtained by SSTA and by screened Hartree-Fock approximation (SFHA), respectively.
Plasma-gain-onset density is also indicated in dotted line.
Inset shows the $n$-dependence of $\mu$ at $k_{\rm B}T/E_{\rm 2D}=0.1$ (solid line) and $0.25$ (broken line).
(b) Contour plot of the common logarithm of the screening parameter $\kappa$.
Results evaluated by our SSTA and by free-carrier theory are presented in solid and chain lines, respectively.
Thick broken line represents the condition, $\kappa_{\rm DH}=\kappa_{\rm TF}$, where $\kappa_{\rm DH}$ and $\kappa_{\rm TF}$ denote Debye-H\"{u}ckel and Thomas-Fermi screening parameters, respectively.
The contour of ionization ratio, $\alpha=0.5$, is also indicated in the dotted line.}
\label{fig4}
\end{figure}
\subsection{Thermodynamic quantities}
Figure \ref{fig4}(a) shows the contour plot of the e-h chemical potential, $\mu=\mu_{\rm e}+\mu_{\rm h}$, on the $n$-$T$ plane.
The solid lines obtained by our SSTA are compared with the broken ones by the screened Hartree-Fock approximation (SHFA)\cite{Haug1985,Schmitt-Rink1985,Schmitt-Rink1986,Ell1989,Ell1990}.
\par
Let us review the SHFA result.
Here, the chemical potentials, $\mu_{\rm e}$ and $\mu_{\rm h}$, and the quasiparticle energies, $\xi_{{\rm e},k}$ and $\xi_{{\rm h},k}$, are determined so as to satisfy
\begin{align}
\xi_{a,k}=&\epsilon_{a,k}+\frac 12\sum_q\left(W_q-V_q\right)-\sum_q W_q f_{\rm F}\left(\xi_{a,k}-\mu_a\right),
\end{align}
and
\begin{equation}
n=\frac 2S\sum_{a,\bm k}f_{\rm F}(\xi_{a,k}-\mu_a), 
\end{equation}
for $a={\rm e}$ and ${\rm h}$, using the screened interaction potential, $W_q$, given by Eq.~(\ref{eq:screen}).
The contour lines show gradual slopes in the classical region ($n\lambda_{\rm T}^2\ll 1$) , but become steep in the quantum region ($n\lambda_{\rm T}^2\gg 1$).
The crossover line between these two regions is given by the plasma-gain-onset condition of Eq.~(\ref{eq:plasma-gain-onset density}), which is shown in the dotted line.
At low temperature, the contour lines become sparse particularly at around the unstable region, reflecting that $\mu$ is insensitive to $n$. 
(Remind here that $(\partial\mu/\partial n)_T$ vanishes at the boundary of the unstable region.)
\par
Now, let us discuss our SSTA result.
At high temperature, the contour lines by SSTA follow those by SHFA.
However, at low temperature, there exists a distinct difference between these two results: some of the contour lines are bundled and flow into the pure Mott transition point indicated by the cross symbol.
This implies that $\mu$ shows a discontinuity at that point.
The complicate $n$-dependence of $\mu$ near the unstable region ($k_{\rm B}T/E_{\rm 2D}=0.1$) can be more clearly seen in the inset of Fig.~\ref{fig4}(a).
As $n$ increases, $\mu$ becomes insensitive to $n$, then suddenly increases and again becomes nearly independent of $n$.
The magnitude of the discontinuity at the pure Mott transition is roughly estimated as $E_{\rm 2D}/2$.
The origin of this value will be discussed in Sec.~\ref{sec:single-particle spectra}.
\par
In Fig.~\ref{fig4}(b), the screening parameter is shown as a contour plot.
Solid lines show our SSTA result.
For comparison, the screening parameter by the free carriers is also shown in chain lines, which can be analytically evaluated as
\begin{equation}
\kappa_{\rm free}=\frac{2e^2}{\varepsilon_{\rm b}\hbar^2}\sum_{a={\rm e,h}}m_a\left[1-\exp\left(-\frac{\pi\hbar^2 n}{m_a k_{\rm B}T}\right)\right],
\end{equation}
and asymptotically approaches the Debye-H\"uckel screening parameter,
\begin{equation}
\kappa_{\rm DH}=\frac{4\pi e^2n}{\varepsilon_{\rm b}k_{\rm B}T}=4\pi n\lambda_{\rm L},
\end{equation}
in the classical limit ($n\lambda_{\rm T}^2\ll 1$), and to the Thomas-Fermi one,
\begin{equation}
\kappa_{\rm TF}=\frac{2e^2M}{\varepsilon_{\rm b}\hbar^2}=\frac 1{a_{\rm 2D}}\frac M{m_{\rm r}},
\end{equation}
in the quantum limit ($n\lambda_{\rm T}^2\gg 1$).
The classical-quantum crossover of the screening parameter is given by the condition, $\kappa_{\rm DH}=\kappa_{\rm TF}$, or equivalently,
\begin{equation}
n\lambda^2_{\rm T}=\frac M{m_{\rm r}}.
\end{equation}
Interestingly enough, this crossover density is always larger than the plasma-gain-onset density of Eq.~(\ref{eq:plasma-gain-onset density}).
This means that the exciton Mott crossover and transition take place in the Debye-H\"uckel screening region, $n\lambda^2_{\rm T}\ll M/{m_{\rm r}}$.
\par
Although our SSTA result is well reproduced by the free-carrier screening parameter in the high ionization regime, $\alpha\gtrsim 0.5$, these two results significantly differ at low ionization regime, $\alpha\lesssim 0.5$.
In fact, the screening parameter obtained by SSTA is much smaller than the free-carrier one in the low ionization regime.
The exciton formation suppresses the screening effect, so that the excitons are self-stabilized in the low-$n$-low-$T$ region.
As a result, some contour lines are pushed to the higher e-h density side, and flow into the pure Mott transition point, indicating the discontinuous change in the screening parameter there.
\subsection{Single-particle spectra}\label{sec:single-particle spectra}
One of the interesting features of our SSTA is that the excitonic effect, or equivalently, $T$-matrix contribution is taken into account in the self-energies, $\Sigma_{\rm e}(k,\omega)$ and $\Sigma_{\rm h}(k,\omega)$.
This fact can be confirmed by investigating the single-particle spectra of the electron and the hole, 
\begin{equation}
A_a(k,\omega)=-2{\rm Im}G_a(k,\omega-\mu_a/\hbar)\ \ \ (a={\rm e},{\rm h}) .
\end{equation}
\par
\begin{figure}[tbp]
 \begin{center}
  \includegraphics[width=0.47\textwidth]{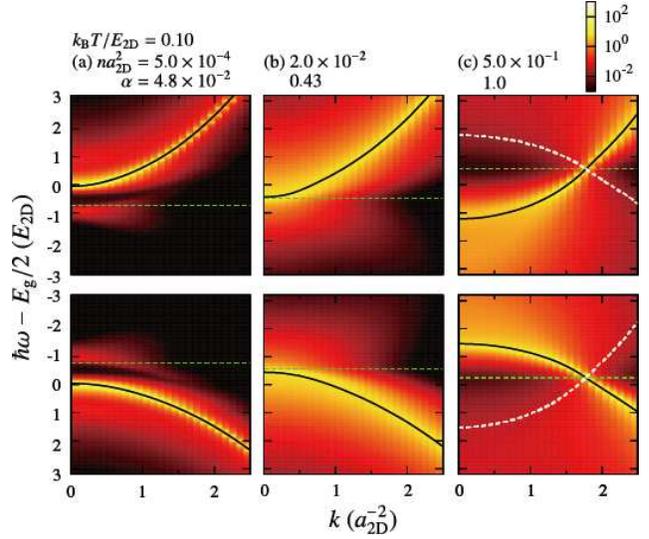}
 \end{center}
 \caption{(color online) Intensity plot of single-particle spectra for electron (upper panel) and hole (lower panel) at a low temperature, $k_{\rm B}T/E_{\rm 2D}=0.1$.
Electron-hole densities are given as (a) $na_{\rm 2D}^2=5\times 10^{-4}$, (b) $2\times 10^{-2}$, and (c) $0.5$.
The quasielectron and quasihole energies, $\xi_{{\rm e},k}$ and $\xi_{{\rm h},k}$, are shown in the black solid lines.
Chemical potentials of the electron and the hole, $\mu_{\rm e}$ and $\mu_{\rm h}$, are also indicated by the broken lines.
The ``shadow'' dispersions of quasihole and quasielectron, $\mu-\xi_{{\rm h},k}$ and $\mu-\xi_{{\rm e},k}$, are represented in the white dotted lines in (c).}
 \label{fig5}
\end{figure}

Figure \ref{fig5} shows the single-particle spectra evaluated at three different e-h densities at low temperature, $k_{\rm B}T/E_{\rm 2D}=0.1$.
The spectra of the electron and hole are shown in the upper and lower panels, respectively.
The quasiparticle energies, $\xi_{a,k}$, are also plotted in the black lines.
They clearly demonstrate that our definition of quasiparticle energies is quite reasonable: the spectra take the maxima in the vicinity of $\hbar\omega=\xi_{a,k}$.
These quasiparticle branches show red shifts as the e-h density increases, indicating the band-gap renormalization. 
In contrast, their masses are almost insensitive to the e-h density.
\par
At the low e-h density, $na_{\rm 2D}^2=5\times 10^{-4}$, the exciton satellite branches are found in addition to the main quasiparticle branches, as shown in Fig.~\ref{fig5}(a).
These satellite branches are located at around $\hbar\omega\sim E_{\rm g}/2-E_{\rm 2D}$ with slightly negative masses , and are ascribed to the processes in which an electron (or a hole) with momentum, $\bm k$, is removed from the exciton, leaving a hole (an electron) with a residual momentum. 
Since the exciton has small center-of-mass momentum at low $T$, the exciton satellite peak of $A_a(k,\omega)$ is found at around
\begin{align}
\hbar\omega&\sim (E_{\rm g}-E_{\rm 2D})-\epsilon_{\bar a,-k}\notag\\
&=\frac{E_{\rm g}}2-E_{\rm 2D}-\frac{\hbar^2 k^2}{2m_{\bar a}},
\end{align}
at fixed $k$, where $\bar a={\rm h}$ and ${\rm e}$ for $a={\rm e}$ and ${\rm h}$, respectively.
It should be noted here that $\bm k$ denotes the momentum of the annihilated particle $a$, so that the momentum of the left particle $\bar a$ is $-\bm k$.
The above result explains the reason why the exciton satellite branch is located at around $\hbar\omega\sim E_{\rm g}/2-E_{\rm 2D}$, and why its dispersion has a negative mass at low temperature.
\par
\par
Fig. \ref{fig5}(b) shows the single-particle spectra near the plasma-gain-onset density, $na_{\rm 2D}^2=0.02$, where the band gap renormalization is comparable to the exciton binding energy, $E_{\rm g}-E_{\rm g}^*\sim E_{\rm 2D}$.
Interestingly enough, we can still find remnants of the exciton satellite branches in the spectra.
While both quasiparticle and exciton satellite branches broaden and show the spectral overlap, the energy of the exciton satellite branch at $\bm k=0$ still stays near $E_{\rm g}/2-E_{\rm 2D}$, which is significantly lower than that of the quasiparticle one, $E_{\rm g}^*/2$.
Owing to this exciton satellite structure, the exciton ionization ratio becomes significantly less than unity ($\alpha=0.43$).
The insensitivity of the energy of the exciton satellite branch is a consequence of the charge neutrality of the e-h pair: the energy of an electron (a hole) correlated strongly with a hole (an electron) is hardly affected by the other electrons and holes in the background.
It is also noteworthy that $\Delta=E_{\rm g}^*/2-(E_{\rm g}/2-E_{\rm 2D})\sim E_{\rm 2D}/2$ can be interpreted as the energy to dissociate such a correlated e-h pair, which explains why the  magnitude of the discontinuity in the e-h chemical potential is approximately equal to $E_{\rm 2D}/2$ at the pure Mott transition point.
\par
At the high e-h density, $na_{\rm 2D}^2=0.5$, spectra show an unusual feature as shown in Fig.~\ref{fig5}(c).
In the upper panel, this feature can be most clearly seen in the ''shadow'' of the hole energy dispersion, $\bar\xi_{{\rm h},k}=\mu-\xi_{{\rm h},k}$, given in the white line. 
Then, we find that the spectral weight of $A_{\rm e}(k,\omega)$ spreads mainly in the energy region expressed by the inequality, $(\hbar\omega-\xi_{{\rm e},k})(\hbar\omega-\bar\xi_{{\rm h},k})>0$.
This redistribution of the spectral weight is understood as the tendency of the mode repulsion near $\hbar\omega\sim\mu_{\rm e}$ between the ``original''and the ``shadow'' dispersions due to the e-h interaction, or equivalently, as a precursor of the e-h Cooper pairing. 
A corresponding feature of $A_{\rm h}(k,\omega)$ is also found in the lower panel.
\par
\begin{figure}
 \begin{center}
  \includegraphics[width=0.47\textwidth]{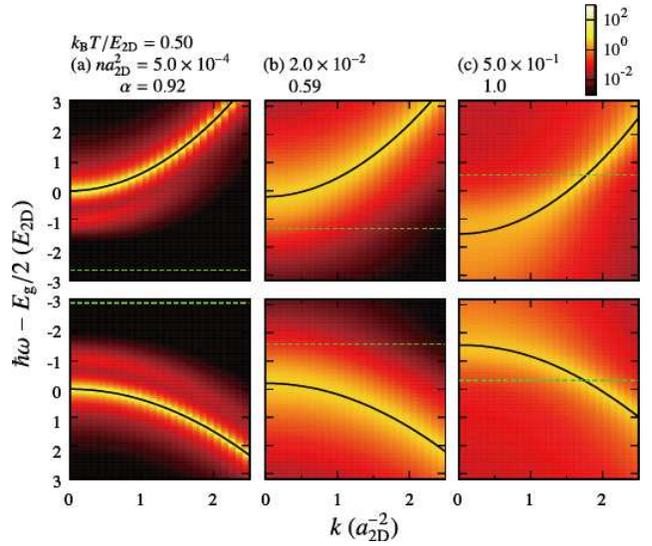}
 \end{center}
 \caption{(color online) Intensity plot of the single-particle spectra at a high temperature, $k_{\rm B}T/E_{\rm 2D}=0.5$.}
 \label{fig6}
\end{figure}
Figure~\ref{fig6} shows the single-particle spectra at higher temperature, $k_{\rm B}T/E_{\rm 2D}=0.5$.
At the low e-h density, $na_{\rm 2D}^2=5\times 10^{-4}$, we again find the exciton satellite branches located at around $\hbar\omega\sim E_{\rm g}/2-E_{\rm 2D}$.
However, their energy dispersions have positive masses, and are almost parallel to the main quasi-particle ones.
At high temperature, the exciton with finite center-of-mass momentum, $\bm Q$, can contribute to the exciton satellite branches, which consists of an electron with momentum near $m_{\rm e}\bm Q/M$ and a hole with momentum near $m_{\rm h}\bm Q/M$.
The exciton satellite branch seen in $A_a(k,\omega)$ arises from the process which annihilates the particle $a$ with momentum, $\bm k\sim m_a\bm Q/M$, leaving the other particle $\bar a$ with momentum, $m_{\bar a}\bm Q/M\sim m_{\bar a}\bm k/m_a$.
Thus, it appears at around
\begin{align}
\hbar\omega
&\sim E_{\rm g}-E_{\rm 2D}+\frac{\hbar^2Q^2}{2M}-\epsilon_{\bar a,m_{\bar a}k/m_a}\notag\\
&\sim E_{\rm g}-E_{\rm 2D}+\frac{\hbar^2(Mk/m_a)^2}{2M}\notag\\
&\hspace{1cm}-\left(\frac{E_{\rm g}}2+\frac{\hbar^2(m_{\bar a}k/m_a)^2k^2}{2m_{\bar a}}\right)\notag\\
&=\frac{E_{\rm g}}2-E_{\rm 2D}+\frac{\hbar^2k^2}{2m_a},
\end{align}
at fixed $k$, which is indeed parallel to the main quasi-particle branch, $\xi_{a,k}\sim E_{\rm g}/2+\hbar^2k^2/2m_a$.
\par
As shown in Fig.~\ref{fig6}(b) and (c), the single-particle spectra are highly broadened at the higher e-h density.
The precursor of the Cooper pairing is not seen even at the highest e-h density.

\section{Summary and Discussion}\label{summary}
We studied exciton-Mott crossovers and transitions in the 2D e-h system by developing a self-consistent screened $T$-matrix approximation, where the self-energy (band renormalization), the screening parameter (interaction renormalization), and the inter-carrier $T$-matrix (excitonic effect), are determined self-consistently.
The characteristics of our theory are that the self-stabilization mechanism of the excitons is taken into account, which is caused by the screening suppression due to the exciton formation.
The phase diagram is illustrated as the contour plot of the exciton ionization ratio in the wide range of the e-h density, $n$, and the temperature, $T$.
It gives detailed information on the Mott physics, beyond the conventional understanding by the Mott density.
\par
While the ionization ratio follows the Saha equation in the classical regime, it rapidly decreases in the quantum regime.
The crossover between these two regimes can be expressed as $n\lambda_{\rm T}^2\sim 0.3$ at high $T$, but deviates from this simple condition at low $T$ due to the self-stabilization of the excitons.
Pure Mott transition point and the region unstable toward the inhomogeneity are also found in our phase diagram at the lower temperature.
These transitions also appear as the singular $n$-dependence of the e-h chemical potential and the screening parameter.
\par
Since the excitonic effect is taken into account in the self-energies, the excitonic satellite branches appear in the single-particle spectra at low $n$, which have negative masses at low $T$ but positive masses at high $T$.
At low-$T$-high-$n$ region, the spectra exhibit the precursor behavior of the e-h Cooper pairing.
\par
It is interesting to compare our present result in the 2D e-h systems with that in the quasi-1D e-h ones\cite{yoshioka2011,yoshioka2012}.
Surprisingly and interestingly enough, most of the Mott physics are universal between quasi-1D and 2D e-h systems.
Still, we can see two important differences.
One is that the self-stabilization mechanism of the excitons is more significant in 2D than in quasi-1D.
The line of the ionization minimum follows the simple classical-quantum crossover line in quasi-1D in the wide range of temperature, while such tendency is restricted only to higher temperatures in 2D.
The other difference is that the ionization ratio at the plasma-gain-onset density is smaller in 2D than in quasi-1D.
The incompleteness of the ionization, $1-\alpha$, is only a few percent even at low $T$ in quasi-1D, but reaches dozen percent in 2D.
These two differences are ascribed to the density of state at the renormalized band edge: it is more enhanced in the lower dimension so that the excitonic correlation is more easily suppressed in quasi-1D than in 2D.
\par
Pure Mott transition point and the unstable region found in our  phase diagram suggest that some intriguing first order transition may take place at low-$T$ region. 
Several possible scenario could be considered; two successive phase transitions accompanied by the phase coexistence, appearance of the novel inhomogeneous phase, and so on.
However, our scheme is still insufficient to discuss this issue, because three- or four-body (trionic or biexcitonic) correlations\cite{Asano2005,Maezono2013} are not fully taken into account.
More elaborate treatment of the Mott phase transitions is our future problem.
\par
In recent experments\cite{Suzuki2012}, the exciton Mott crossover is investigated by means of optical pump and terahertz probe spectroscopy.
It is not certain whether the ionization ratio estimated from such measurements directly corresponds to the one employed in our study.
Still, we expect our phase diagram to provide indispensable information for the development to these experimental studies\cite{note}.

\begin{acknowledgments}
This work is supported by KAKENHI (Nos 20104010, 21740231 and 25400327).

\end{acknowledgments}


\begin{thebibliography}{00}
\bibitem{correction}
In this sense, the contour lines of $\alpha=1$ should be removed from Fig.~2 in Ref.~\citen{yoshioka2011} and from Fig.~2 in Ref.~\citen{yoshioka2012}, which merely indicate that $\alpha$ can be regarded as unity within the numerical error.
\end{thebibliography}
\end{document}